\documentclass[pra,superscriptaddress,,nofootinbib]{revtex4}
\usepackage{graphicx}
\usepackage{amsmath}
\usepackage{amssymb}
\usepackage{float}
\usepackage{algorithm}
\usepackage{amsthm}

\floatname{algorithm}{Protocol}
\usepackage{bm}
\usepackage{bbm}

\usepackage{color}

\def\EQ#1{\begin{eqnarray}#1\end{eqnarray}}
\def\Tr{{\mathrm{Tr}}}
\newcommand{\djj}{d\kern-0.4em\char"16\kern-0.1em}
\newtheorem{theorem}{Theorem}
\newtheorem{lemma}{Lemma}

\newtheorem{prop}{Proposition}\def\PRO{\begin{prop}}\def\ORP{\end{prop}}
\newtheorem{coro}{Corollary}\def\COR{\begin{coro}}\def\ROC{\end{coro}}
\newtheorem{theo}{Theorem}\def\TH{\begin{theo}}\def\HT{\end{theo}}
\def\TH{\begin{theo}}\def\HT{\end{theo}}
\newtheorem{defi}[prop]{Definition}\def\DE{\begin{defi}}\def\ED{\end{defi}}
\newtheorem{lemme}[prop]{Lemma}\def\LE{\begin{lemme}}\def\EL{\end{lemme}}
\def\ket#1{\left| #1 \right\rangle}
\def\bra#1{\left\langle #1 \right|}

\begin{document}
\title{Quantum digital signatures with quantum key distribution components}

\author{Petros Wallden\footnote{V. Dunjko and P. Wallden contributed equally to this work.}\footnote{petros.wallden@gmail.com}}
\affiliation{LFCS, School of Informatics, University of Edinburgh, 10 Crichton Street, Edinburgh EH8 9AB, UK.}
\affiliation{Institute of Photonics and Quantum Sciences, School of Engineering and Physical Sciences, David Brewster Building,  Heriot-Watt University, Edinburgh EH14 4AS, UK.}
\author{Vedran Dunjko
}
\affiliation{Institute for Quantum Optics and Quantum Information,
Austrian Academy of Sciences, Technikerstr. 21A, A-6020 Innsbruck, Austria}
\affiliation{Institut for Theoretical Physics, University of Innsbruck, Technikerstra{\ss}e 25, A-6020 Innsbruck, Austria}
\affiliation{Laboratory of Evolutionary Genetics, Division of Molecular Biology, Ru\djj er Bo\v{s}kovi\'{c} Institute, Bijeni\v{c}ka cesta 54, 10000 Zagreb, Croatia.}

\author{Adrian Kent
}
\affiliation{Centre for Quantum Information and Foundations, DAMTP,
Centre for Mathematical Sciences, University of Cambridge,
Wilberforce Road, Cambridge, CB3 0WA, U.K}
\affiliation{Perimeter Institute for Theoretical Physics, 31 Caroline Street North, Waterloo, ON N2L 2Y5, Canada}

\author{Erika Andersson
}
\affiliation{SUPA, Institute of Photonics and Quantum Sciences, School of Engineering and Physical Sciences, David Brewster Building, Heriot-Watt University, Edinburgh EH14 4AS, UK.}

\begin{abstract}
Digital signatures guarantee the authenticity and transferability of messages, and are
widely used in modern communication. The security of currently used classical digital signature schemes, however, relies on computational assumptions.
In contrast, quantum digital signature (QDS) schemes offer information-theoretic security guaranteed by the laws of quantum mechanics.
We present two
QDS protocols which have the same experimental requirements as quantum key distribution,
which is already commercially available.
We also present the first security proof for any QDS scheme
against coherent forging attacks.
\end{abstract}

\maketitle

\section{Introduction}

Digital signatures are
commonly used to guarantee the identity of a sender and the
authenticity
of a
message, for example in electronic commerce and e-mail. Importantly, digital signatures also guarantee that messages are transferable, so that
a forwarded message will also be accepted as valid.
These cryptographic tasks are different from ensuring that a message
is kept secret. Rivest, one of the inventors of the RSA algorithm for public key cryptography, wrote in 1990 that ``the notion of a digital signature may prove to be one of the most fundamental and useful inventions of modern cryptography"~\cite{Rivest1990}.
Currently used classical digital signature schemes employ
public key encryption,
where security relies on conjectured but unproven computational hardness of cryptographic functions.
In contrast, quantum digital signature schemes \cite{QDS,ErikaOrig,OurNatComm,QDS_prl,QDS_exp}, which are
quantum versions of Lamport's one-time
signature scheme
~\cite{Lamport}, offer information-theoretic security relying on the laws of quantum mechanics.

We mainly consider the simplest non-trivial setting for digital signatures,
with three parties, which is sufficient to illustrate how our
protocols work.
Alice
signs the message, Bob
first receives the message and needs to authenticate it, and Charlie
receives the forwarded message from Bob, and
verifies that the initial source was indeed Alice. The desired protocol needs to be secure against cheating,
provided that at most one of the three parties is dishonest. We require security both against message forging by Bob and against
repudiation by Alice
\footnote{These are the most significant forms of cheating. Our protocols can
be extended to the many-party setting, and to deal with general cheating attacks in that context, but since this complicates
the protocols and security analysis somewhat we postpone full discussion for future work.
}.
In our setting, successful repudiation by Alice means that a
message is accepted by Bob but would be rejected when forwarded to
Charlie, that is, the message is not transferable
\footnote{Repudiation by Alice means that she can successfully    deny having sent a message that she actually did send. Preventing repudiation is    closely related to, but not in general equivalent to,
    ensuring message    transferrability, i.e. ensuring that forwarded messages are accepted as valid.
    For example, a poorly designed protocol can fail to
    ensure transferability even if Alice is honest.
   In our scenario with one sender and two
    receivers,
    non-repudiation and message transferrability become
    identical if a majority decision is used to resolve disputes.}.
In our QDS
protocols, it is easier to forge a message when claiming it to
be forwarded, and in forging scenarios we therefore assume that Bob is
trying to forge a forwarded message.

QDS schemes have two stages, the distribution stage and the messaging stage. In the latter stage, a message is actually sent and signed. While details vary,
different schemes
share common features. During the distribution stage Alice sends a
quantum signature, a sequence of quantum states, to Bob and Charlie.
In order to prevent
repudiation, they can either
compare their states \cite{QDS} or symmetrise them
\cite{ErikaOrig,OurNatComm,QDS_prl,QDS_exp}. Bob and Charlie then either store the quantum signature, or
measure it and store the outcomes. In the messaging stage, which could occur much later, Alice wants to sign a message. During the messaging stage, Alice sends the classical description of the quantum signature, and Bob and Charlie
confirm that this is compatible with their stored information. Importantly, the participants must be able to decide on the validity and transferrability of the message without further communication with other participants.

The first QDS protocol, proposed by Gottesman and Chuang \cite{QDS}, required
 involved processing of the quantum signatures -- a general SWAP test
 and long-term quantum memory --
 which is currently unfeasible experimentally.
In
\cite{ErikaOrig}, an optical multiport
replaced the
SWAP test. Long-term quantum memory was however still required.
To remedy this,
we suggested a
protocol~\cite{QDS_prl} where the signature states are measured directly in
the distribution stage. This protocol guaranteed security against collective attacks, but still employed a multiport to ensure non-repudiation.
When implemented~\cite{QDS_exp}, however,
the multiport caused substantial losses. Aligning the multiport becomes  increasingly difficult when the distance between recipients increases.
Here we therefore propose two schemes for quantum digital signatures which require neither quantum memory, nor a multiport. They require only the same components as quantum key distribution (QKD), enabling existing QKD ``hardware" to be used also for QDS. This significantly extends and enhances the use of QKD systems.

The two protocols are denoted P1 and P2.
There are many possible variations on these protocols, e.g. using different
quantum states
for the quantum signatures (such as
phase-encoded coherent states) or different types of measurements (unambiguous quantum state discrimination,
minimum-error measurements, etc.).
Here, we will focus on protocols employing BB84 states, as they are well-studied in the context of QKD, and this choice allows for the first proof of security against coherent forging attacks.

P1 is inspired by the protocol in~\cite{QDS_prl,QDS_exp},
while P2
only uses quantum-mechanical features
to produce secret shared classical keys using QKD. After this,
P2 continues with
a classical scheme, with information-theoretic security
relying on the security of the shared secret keys.
This means that the functionality of information-theoretically secure digital signatures follows directly
from point-to-point QKD. To our knowledge, there are few
information-theoretically secure classical digital signature schemes
based on secret shared keys, and all of them require extra assumptions such as the existence of a trusted third party \cite{Hanaoka2000,SwSt2011} or the existence of authenticated broadcast channels \cite{CR1991}.
Using P2, the functionality of digital signatures is implied by
sharing secret keys alone.
Since P1 uses the same ``quantum hardware" as the  generation of secret keys by QKD, for use in P2, it is an open question whether P1 or P2 is most resource-efficient, in particular when generalising to more than three parties.

Just as for QKD, for QDS schemes one assumes that between all parties, Alice, Bob and Charlie, there exist pairwise authenticated classical
channels, guaranteeing that classical messages cannot be tampered with.
Such channels are resource-inexpensive \cite{MAC}.
Moreover, for both QKD and QDS it is essential that participants
can be sufficiently sure that if
a quantum state is sent,
then (approximately) that same quantum state is also received,
without an eavesdropper or forger having learnt (too much) about it.
How to achieve this is well established
for QKD, and we expect that similar techniques can be used for QDS. We further comment on this
in the discussion at the end. However, for the moment,
we will for P1 make the stronger assumption (which existing QDS
protocols also make) that there are authenticated ideal
quantum channels between the participants. This guarantees that the quantum state any participant
sends is received by the intended recipient.
Nevertheless, we formulate our protocol with non-ideal channels in mind, and also note that analysis of previous QDS experiments~\cite{OurNatComm, QDS_exp} has considered imperfections in scenarios with only individual forging attacks.

\section{Protocol P1}

A main difference between P1 and the
protocols in \cite{QDS_prl,QDS_exp}
is that a multiport is not needed. Instead, security against repudiation is guaranteed by Bob and Charlie exchanging some of their signature elements, leading to a significantly simpler experimental implementation.
In the basic version of P1, the exchange is done before measuring the signature states, and in a modified version P1', described in Appendix~\ref{Supp}, after measuring them.
We will use the same four quantum states as the BB84 protocol for quantum key distribution~\cite{BB84},
given by
\begin{equation}
|0\rangle,~~ |1\rangle,~~ |+\rangle=\frac{1}{\sqrt{2}}(|0\rangle+|1\rangle),\ \textup{and}~~|-\rangle=\frac{1}{\sqrt{2}}(|0\rangle-|1\rangle).
\end{equation}
As discussed above, we assume that between all parties, Alice, Bob and Charlie, there exist authenticated classical and quantum channels.

\noindent\textit{Distribution stage}

\noindent 1. For each possible future one-bit message $k=0,1$, Alice generates
two copies of sequences of BB84 states,
$QuantSig_k = \bigotimes_{l=1}^L \rho^k_l$ where
$\rho^k_l$ is a randomly chosen BB84 state, $\ket{\psi^k_l}=\ket{b^k_l}$, and $b^k_l\in\{-,0,+,1\}$, and $L$ is a suitably chosen integer.
The state $QuantSig_k$ and the sequence of signs $PrivKey_k = (b_1^k, \ldots b_L^k)$ are called the \textit{quantum signature} and the \textit{private key}, respectively, for message $k$.
The individual state $\rho_{l}^k$ we call
the $l^{th}$ \emph{quantum signature element} state for
message $k$.

\noindent 2. Alice sends one copy of $QuantSig_k$ to Bob and one to Charlie,
for each possible message $k=0$ and $k=1$.

\noindent 3. Bob (Charlie), for each element $l$ of
$QuantSig_k$ for $k=0,1$, randomly chooses
to either
forward the signature element to Charlie (Bob), or keep it and directly measure it as described under 4. below. In either case,
the position $l$ is recorded.
We should note here that it is not important that Bob and Charlie exchange states at the same time. The protocol is secure even if the
signature element exchanges are not synchronized. This is a significant improvement over multiport-based schemes where near-perfect synchronization was essential \footnote{Exchanging parts of the signature in practical implementations leads to losses increasing with the distance between the recipients. However, the multiport used in e.g. \cite{QDS_prl} also incurs other substantial losses. In addition to losses in multiport beam splitters and other optical elements, a more serious issue is the increasingly difficult synchronization between Bob and Charlie, since even a slight time shift significantly decreases visibility.}.

\noindent 4. Bob (Charlie) measures the states he kept and the states that Charlie (Bob) sent him, randomly choosing either the $\{\ket{0},\ket{1}\}$ basis or the $\{\ket{-},\ket{+}\}$ basis. In this way, for each signature element Bob or Charlie measures, each of them unambiguously excludes one of the four possible states. For example, if Bob obtains the measurement result ``$|1\rangle$", this means
that Alice cannot have sent the state $|0\rangle$.
Bob and Charlie record what state they excluded, for each element $l$ and message $k$. This type of quantum measurement
is called quantum state elimination~\cite{stevebook,OppenUSE,QDS_exp}. The sequence of excluded states
will later be used to authenticate or verify a message. We call this an \emph{eliminated signature}.

\noindent 5. If either Bob or Charlie receives, from the other party,  fewer than $L(1/2-r)$ or more than $L(1/2+r)$ signature elements per possible message,  then they abort.
That is, in the ideal case with no transmission
losses~\footnote{For simplicity, we
assume no transmission losses during this stage. Nevertheless, also for an imperfect realization it should be possible to modify
protocol parameters to ensure security,
following
approaches in \cite{OurNatComm,CK2012,bcexpt}.}, Bob expects on average $L/2$ signature elements from Charlie, and aborts if he receives too few or too many by setting a threshold $r$.
If all participants are honest, then the probability for abort depends on the ``coin" that Bob (Charlie) tosses to decide whether to keep or forward a qubit.
Since the choice is done independently, with equal probabilities for each instance,
it follows that this probability decays exponentially as $L$ increases.

At this point,
for some positions $l$ in the quantum signature, Bob (Charlie) has measured both copies of signature elements which Alice sent,
for some he has measured the signature element copy sent directly to him by Alice,
for some the
copy originally sent to Charlie (Bob), and for some positions he has measured no
copy at all.
Each of these possibilities occurs for on average $L/4$
positions.
Bob, for each signature  element position, has therefore ruled out one, two or none of the four possible states, and similarly for Charlie. These records form Bob's and Charlie's eliminated signatures.

\noindent\textit{Messaging stage}

\noindent 1. To send a signed one-bit message $m$, Alice sends $(m,{\rm PrivKey}_m)$ to the desired recipient (say Bob).

\noindent 2. Bob checks whether $(m,{\rm PrivKey}_m)$ matches his stored
eliminated signature by counting how many elements of Alice's private key he actually ruled out in the distribution stage.
If there are fewer than $s_a L$ mismatches , where $s_a$ is a small \textbf{authentication threshold} (zero in the ideal case), Bob accepts the message.

\noindent 3. To forward the message to Charlie, Bob forwards to Charlie the pair  $(m, {\rm PrivKey}_m)$ he received from Alice.

\noindent 4.
Charlie tests for mismatches similarly to Bob, but in order to protect against repudiation by Alice, the threshold is different. Charlie accepts the forwarded message if there are fewer than $s_v L$ mismatches, where $s_v$ is the \textbf{verification threshold}, with $0 \leq s_a < s_v <1$.

\emph{Security analysis}. Digital signature schemes should be secure against
both repudiation and forging. Security against repudiation guarantees that Alice cannot make Bob and Charlie disagree on the validity
(and consequently the content) of her message (except with very small probability).
Security against forging means that any recipient will with high
probability reject any message which was not originally sent by Alice
herself. The security analysis is
outlined below, with more details
in Appendix~\ref{Supp}.

Security against repudiation: Alice initially sends (possibly different) strings of BB84 states to Bob and Charlie. More generally, she could send any states, including entangled states. Bob and Charlie randomly choose to keep or forward each of the signature elements. From Alice's perspective,
at the end of the distribution stage, the reduced density matrices for
Bob's and Charlie's quantum states are identical, irrespective of what
states she sent them. Intuitively (see Appendix~\ref{Supp} for a proof),
she thus has little chance of
making Bob accept and Charlie reject the same declaration.
 Moreover, Alice gains nothing by sending different quantum signatures to Bob and Charlie. Her best strategy
is to
send a declaration with $L(s_v-s_a)/2$ mismatches with
the quantum signature she sent.
Her probability for repudiation is then
\EQ{p({\rm rep})\leq\exp[-(s_v-s_a)^2L/2]}
which, since $s_a<s_v$, decays exponentially as the length $L$ of the signature increases.
Note that setting a non-zero $s_a$ will be necessary if the quantum channels are not ideal.

Security against forging: In order to successfully forge, Bob needs to
guess, causing fewer than $Ls_v$ mismatches, the part of the signature that Alice sent to Charlie and which Charlie kept. In so-called individual
forging attacks, Bob makes measurements on individual signature elements.
Bob, in order to make the best possible guess, should then perform minimum-error measurements on his elements.
One can show (Appendix~\ref{Supp},\cite{WDA2013}) that for each
element, the minimum probability for Bob to declare a mismatch is $1/8$, leading to a bound
\EQ{ \label{indBound}p({\rm forge})\leq\exp[-2(1/8-s_vL/K)^2K]}
on the forging probability,
where $K=L(1/2-r)$ is the number of elements that Charlie kept.  This probability decays exponentially with the length $L$ of the signature provided that $s_v<1/8 (K/L)=1/16(1-2r)$.
In fact, the bound in Eq. (\ref{indBound}) is the best a forger can
achieve with any strategy, including coherent attacks.
To show this, we follow the arguments of \cite{CK2012} for the
security of a relativistic quantum bit commitment protocol
\cite{Kent2012}.
The central result we employ shows that no coherent measurement strategy can beat a local strategy in correctly declaring
the state of an individual signature key element, even if one post-selects on any measurement outcomes for all other elements. We can then show that the individual strategies for forging are optimal: see Appendix\ref{Supp} for details.  Note that this proof applies
specifically to the BB84 versions of the protocols considered here, and for example does not generalize to versions considered previously using B92 states.

\section{Protocol P2}

The second protocol, P2, achieves the functionality of QDS by using only (long) shared keys and untrusted classical channels. Shared keys can, of course, be achieved using a secure classical channel.
Alternatively, QKD can be used for generating shared keys, with information-theoretic security. If QKD is thought of as key expansion, this requires only short pre-shared keys, effectively independent of future message size.

For QKD, we must also
assume that untrusted quantum channels are available.
In short, protocol P2
may be based on point-to-point QKD, which is under development in many research groups and even commercially available~\cite{idquantique,magiq,sequrenet,anhui,quintessence}.

\noindent\textit{Distribution stage}

\noindent 1. For each possible future message $k=0,1$, Alice generates two \emph{different} secret keys (called \emph{signatures}) consisting of sequences of classical bits. We call an individual bit the $l^{th}$ \emph{signature element} for message $k$.

\noindent 2. For each possible message $k=0,1$, Alice sends one secret key to Bob and the other to Charlie via secure classical channels.

\noindent 3. For each signature element
and for $k=0,1$, Bob (Charlie) randomly chooses to either keep it
or send it to Charlie (Bob) via a secure classical channel.

\noindent 4. If either Bob or Charlie receives  fewer than $L(1/2-r)$ or more than $L(1/2+r)$ signature elements per possible message from the other party, then the protocol is aborted.

\noindent\textit{Messaging stage}

\noindent 1. To send a signed one-bit message $m$, Alice sends $(m,{\rm PrivKey}B_m,{\rm PrivKey}C_m)$ to the desired recipient (say Bob). That is, Alice declares both private keys corresponding to the message $m$ in order to sign.

\noindent 2. Bob checks whether the declaration
$(m,{\rm PrivKey}B_m,{\rm PrivKey}C_m)$ matches his key and the parts of the key that Charlie sent him. If it does, then he accepts the message. For
classical keys, we can
assume that if all parties are honest then there are no mismatches, and therefore we can
set $s_a=0$.

\noindent 3. To forward the message to Charlie, Bob forwards to Charlie  the declaration $(m, {\rm PrivKey}B_m,{\rm PrivKey}C_m)$ he received from Alice. Charlie tests for mismatches similarly to Bob, but
accepts the forwarded message if the following two conditions are met. (i) There is no mismatch between the declaration
and the part of ${\rm PrivKey}B_m$ which Charlie obtained from Bob and (ii) there are fewer than $s_vL$ mismatches between the  declaration
and Charlie's ${\rm PrivKey}C_m$, where the verification threshold
for security against repudiation satisfies $1/2>s_v>0$.

Security against repudiation: Alice needs to make Bob accept the message while Charlie rejects. This means that Alice's declaration cannot have any mismatch with Bob's key, and necessarily at least $s_vL$ mismatches with Charlie's key.
The probability for repudiation then satisfies
\EQ{p({\rm rep})\leq (1/2)^{s_vL},}
where the RHS decays exponentially with increasing signature length $L$ (more details in Appendix~\ref{Supp}).

Security against forging: Bob needs to guess, with fewer than $Ls_v$ mismatches, the $K\geq L(1/2-r)$ elements of Charlie's key that he did not receive (i.e. provided no abort occurred). The probability for each correct guess is $1/2$, and the forging probability therefore satisfies
\EQ{p({\rm forge})\leq \exp\left[-4\left(1/4-\frac{s_v}{1-2r}\right)^2L(1-2r)\right].}
Provided that $s_v<1/4-r/2$, this decays exponentially with increasing
$L$ (more details in Appendix~\ref{Supp}).

\section{Discussion}
We have here proposed and examined QDS schemes suitable for implementation
with current technology, in particular, with the same requirements as for QKD systems. In previous schemes~\cite{QDS_prl, QDS_exp}, while the very demanding requirement for quantum memory was removed, transferrability was guaranteed using a multiport. The multiport leads to
high losses and greater experimental complexity, severely restricting the distance between Bob and Charlie.
To obtain a truly feasible QDS scheme we here suggested two (main) QDS
protocols that do not require a multiport. Protocol P1, other than the
multiport, requires similar resources as protocols in~\cite{QDS_prl,
  QDS_exp}. Importantly, the simplifications we have introduced also
allow a security proof of QDS against coherent forging attacks.
For protocol P2,  we suggest that QKD is used to obtain classical secret keys shared pairwise between all parties. The long shared keys are then shown to
enable the functionality of (Q)DS.
P2 is,
to our knowledge, is the first information-theoretically secure classical digital signature scheme relying only on secret shared keys,
without further assumptions such as a trusted third party or authenticated broadcast channels \cite{Hanaoka2000,SwSt2011,CR1991}.
This illustrates how novel classical protocols can arise inspired by quantum information science.

We now briefly address the question of how one could relax the
assumption of quantum authenticated channels,
while still preventing man-in-the-middle attacks
and other eavesdropping attacks. It is likely that one could use procedures analogous to the
parameter estimation (PE) phase of QKD protocols. Here,
Alice and Bob sacrifice a random selection of quantum states in order to establish how correlated their measurement outcomes are.
Based on the level of correlations in the announced bits, they can deduce that the remaining (unannounced) bits are similarly correlated,
using the quantum de Finetti theorem \cite{Renner05,Renner07}. The protocol is aborted if the level of correlations is insufficient.
A similar approach could be used in QDS: Alice and each of the recipients could in the distribution stage sacrifice parts of the quantum signatures. The level of correlations could be used to infer the level of correlations between Alice's private key and the classical measurement outcomes the recipients obtain in the distribution stage, in analogy with PE. Again, a suitable threshold (related to the signature length $L$, or more precisely, the desired security level) on the correlations should be imposed, and the protocol should be aborted if it is violated. We note that in QDS,
since the participants
need not---and should not!---have identical signatures, other types of classical post-processing used in QKD, such as information reconciliation, may not be required. We leave a full rigorous investigation of this for future work.

Many other open questions still remain.
For instance, a composable security analysis for both protocols is still an open issue.
It is also important to examine exactly how to generalise the presented protocols for more than three parties, or for signing longer messages. For instance, one needs to allow for coalitions of malevolent participants.
Finally, entanglement-based protocols which may lead to device-independent QDS can also be envisaged.

\acknowledgments
 V.D. thanks Dominique Unruh for discussions that eventually led to ideas for protocol P2. This work was supported by the UK Engineering and Physical Sciences Research Council (EPSRC) through grants EP/G009821/1, EP/K022717/1, and the EPSRC Doctoral Prize fellowship grant. P.W. gratefully acknowledges partial support from COST Action MP1006. AK was partially supported by a grant from FQXi and by Perimeter Institute for Theoretical Physics. Research at Perimeter
Institute is supported by the Government of Canada through Industry Canada and by the Province of Ontario through the Ministry of Research and Innovation.

\appendix

\section{Supplementary Material}\label{Supp}

We first give a protocol P1', which is similar to protocol P1 but assumes somewhat different resources. Then we
discuss the security of protocol P1 against both repudiation and forging. Finally we examine the security of protocol P2.

\subsection{Modified protocol P1'}
Here we outline a
modified version of P1 which we call P1'. While the security analysis of P1' is essentially identical to that of P1, P1' uses different resources. In particular, the assumption of an authenticated quantum channel between Bob and Charlie is replaced by the assumption of a secure classical channel between Bob and Charlie (which could be, for instance, achieved by using QKD and an authenticated classical channel). The changes in the protocol are the following:

\begin{itemize}

\item
When Bob (Charlie) receives the quantum signature from Alice, he immediately measures all qubits he receives, using the same measurement as in P1 to exclude one of the possible states. We refer to this as an unambiguous state elimination (USE) measurement.

\item
Subsequently, Bob (Charlie) for each element of the signature randomly decides to (i) either keep the outcome in classical memory or (ii) send the outcome via a classical secure channel to Charlie (Bob). In the latter case, they
will not use the classical record of the outcome in the subsequent protocol (if they are honest -- this is to make things fully symmetric from Alice's point of view).
\end{itemize}

If Bob and Charlie are honest, they end up with precisely the same set of outcomes as they would in protocol P1. The security analysis with respect to repudiation is therefore
identical for P1 and P1'.
If Bob is dishonest, then the security of  P1' is guaranteed by that half of signature elements which Alice sent to Charlie, for which Charlie kept the outcomes. The security analysis for forging is therefore also
identical for protocols P1 and P1'.

\subsection{Security of protocol P1}
The notion of security for QDS is different than for QKD.
For QDS, one needs to separately consider the probability for repudiation (when Alice is malevolent) and forging (when Bob is malevolent). For a protocol to be secure, one requires that both of these probabilities should decay exponentially with the length of the signature $L$.
This implies that any desired level of security $\epsilon$ can be
achieved, while inducing only a logarithmic overhead, by choosing $L$
to be $O(\log (1/\epsilon))$.
Then, one can choose the parameters of the protocol $s_a$ and $s_v$ so as to minimize the maximum overall probability for malevolent behaviour. Typically, this happens when repudiation and forging probabilities are equal, if security against repudiation and forging are considered equally important. Here we will first show that the probability of repudiation decreases exponentially with $L$, and then show the corresponding result for the forging probability. In this paper we assume that quantum authenticated channels are used during the distribution of quantum signatures. Given this assumption, any level of security can be efficiently achieved, irrespective of the level of losses. We note that, however, if one wishes to remove this assumption, then, similarly to QKD protocols, there will be limits on the allowable losses, above which a QDS protocol is no longer secure.

\subsubsection{Security against repudiation}

During the distribution stage, Alice sends $L$ qubits to Bob and $L$ qubits to Charlie for each possible message. To specify which qubit we refer to, we say that qubit $b_i$ is the $i^{th}$ qubit sent to Bob, while $c_i$ the $i^{th}$ qubit sent to Charlie.
Note that during the distribution stage Bob and Charlie exchange qubits, and that  the labels above refer to which person Alice initially sent the qubit to.

At the end of the distribution stage, Bob and Charlie have measured all the $2L$ qubits using USE measurements.
Since we assume that there is an authenticated quantum channel between Bob and Charlie, Alice cannot tamper with the states forwarded from Bob to Charlie and vice versa. From her point of view, each qubit is equally likely to end up being measured by either Bob or Charlie.
For each of the $2L$ qubits, either Bob or Charlie
has ruled out one possible state (out of four BB84 states).
If Alice tries to repudiate a message, she
sends a declaration which she wants Bob to accept and Charlie to reject. For each qubit the declaration either is compatible (a match, which we denote as 1), or is not compatible (a mismatch, which  we denote as 0) with the classically stored
information of what states have been ruled out. We therefore have a sequence of binary outcomes $r=(b_{1},\cdots,b_{L},c_{1},\cdots,c_{L})$ where $b$ and $c$ take values $\{0,1\}$
and $b,c$ refer to the which party the qubit was initially sent while
the subscript denotes the position in the signature the qubit
had. There are $2^{2L}$ different sequences $r$ but not all of them
can be achieved by Alice (e.g. if the state ruled out for $b_m$ and
$c_m$ is different, it is not possible that both $b_{m}$ and $c_{m}$
give a mismatch).

For any fixed sequence of outcomes $r$, there is some probability $p_{\rm rep}(r)$ that Alice repudiates. This depends on which elements end up in Bob's and which end up in Charlie's possession, which is not determined by Alice. By sending the overall quantum signature $\rho_{bc}$ to Bob and Charlie, Alice generates a probability distribution on different outcomes $r$. We will denote the probability of getting outcome $r$ if Alice sends the overall state $\rho_{bc}$ as $p^{\rho_{bc}}(r)$. It follows that the overall repudiation probability given that Alice sends a total state $\rho_{bc}$ is

\EQ{p^{\rho_{bc}}_{\rm rep}=\sum_r p^{\rho_{bc}}(r)\times p_{\rm rep}(r).}
We can see that the probability of repudiation is bounded by $\max_r p_{\rm rep}(r)$. In what follows, we show that $\max_r p_{\rm rep}(r)$ decays exponentially as the length of the signature $L$ increases.

Now we separately consider the subset initially sent to Bob and the subset initially sent to Charlie. Let $\bar p^B_0(r)$ be the average number of mismatches divided by $L$, for
Bob's subset of signature elements in the outcome sequence $r$, and similarly $\bar p^C_0(r)$ for Charlie's subset. That is, for
$r=(b_1,\cdots,b_L,c_1,\cdots,c_L)$, we have
$\bar p^B_0(r)=1-1/L\sum_k b_k$ and $\bar p^C_0(r)=1-1/L\sum_k c_k$.

After randomly exchanging subsystems, the expected proportion
of mismatches for both Bob and Charlie per signature element is the same and is given by
\EQ{p^{B'}_0=p^{C'}_0=1/2(\bar p^B_0+\bar p^C_0)=p,}
where we have suppressed the $r$-dependence for clarity.
We can now see using the Hoeffding
inequality \cite{Hoeff}, that if $p^{B'}_0>s_a$, then the probability of Bob accepting is bounded by
\EQ{p(\textrm{Bob accepting})=P(p-  X^B  \geq p-s_a)\leq \exp(-2(p-s_a)^2L)}
and the probability of Charlie rejecting, provided that $p<s_v$, is
\EQ{p(\textrm{Charlie rejecting})=P(  X^C -p\geq s_v-p)\leq
  \exp(-2(s_v-p)^2L).}
Here $X^B$ and $X^C$ are the actual proportions of mismatches obtained
by Bob and Charlie respectively.
We note that
\EQ{p_{\rm rep}\leq \min\{p(\textrm{Bob accepting}),p(\textrm{Charlie rejecting})\}.}
It follows that the optimum choice for Alice to maximize the probability for repudiation is to aim for $p=(s_a+s_v)/2$, leading to the best repudiation probability
\EQ{p^{\rho_{bc}}_{\rm rep}\leq\exp(-(s_v-s_a)^2L/2)}
which, given that $s_a<s_v$, decays exponentially as the length $L$ of the signature increases. Note that in the main paper, we have for clarity used the simpler notation $p(rep)$ instead of $p^{\rho_{bc}}_{\rm rep}$.

\subsubsection{Security against forging}

The proof follows the following structure. First we derive the best measurement that minimises the probability of a mismatch between the forger's declaration and the honest recipient's measurement, for a single element of the signature. Using this we give a bound on the forging probability if one restricts Bob to measuring each quantum signature element individually. Then we prove that performing a coherent attack, and then conditioning on any sequence of outcomes, cannot increase the success probability for avoiding mismatch for the $N$th element. We prove that this requirement implies that no coherent attack can perform better than the individual attack described earlier. In this proof we follow closely the security analysis of Croke and Kent \cite{CK2012} for the security of the relativistic bit commitment scheme by Kent \cite{Kent2012}. The underlying mathematical problem from the forger's point of view is very similar. Our analysis holds in the case that the protocol is performed using the BB84 states, and we will comment on this at the end.\\

\noindent\emph{Individual Attacks}:

{\lemma\label{lemma1} Suppose Alice selects a single BB84 state $\ket{\psi_A}$, chosen  uniformly at random, prepares two copies of it, and gives one to Bob and one to Charlie. Charlie makes a USE measurement, ruling out one of the three states that Alice did not send, $\ket{\psi_C}$. Then, whatever measurement Bob performs on his copy, the probability $p$ of Bob declaring a single state $\ket{\psi_B}$, which happens to be the one that Charlie ruled out (so $\ket{\psi_B} = \ket{\psi_C}$), is at least $p\geq C_{\rm min}=1/8$.

An optimal strategy that realizes this bound is to measure either in the $\{\ket{0},\ket{1}\}$ or in the $\{\ket{+},\ket{-}\}$ basis, or to perform any POVM $\Pi$ whose elements are weighted combinations of these projective measurements, with measurement operators $\Pi=\{q\ket{0}\bra{0},(1-q)\ket{+}\bra{+},q\ket{1}\bra{1},(1-q)\ket{-}\bra{-}\}$.}

\begin{proof} Finding the minimum probability that Bob's guess is not ruled out by Charlie is a minimum-cost problem. If Bob could always guess what state Alice sent, then he would never generate a mismatch. However, not all mistakes Bob makes will be detected by Charlie with equal probability. If Alice sends the state $\ket{0}$, then it is more likely that Charlie will rule out the state $\ket{1}$ than either of the states $\ket{+},\ket{-}$, so the ``cost'' for Bob making the declaration $\ket{1}$ is greater. The relevant cost matrix is then given by
\EQ{C=\left( \begin{array}{cccc}
0 & 1/4 & 1/2 & 1/4\\
1/4 & 0 & 1/4 & 1/2 \\
1/2 & 1/4 & 0 & 1/4\\
1/4 & 1/2 & 1/4 & 0\end{array} \right),}
where the states appear in the order $(\ket{0},\ket{+},\ket{1},\ket{-})$, and the rows correspond to the state that Alice sent and the columns to the state Bob declares.

One can see that an optimal measurement that Bob can perform is to measure either in the $\{\ket{0},\ket{1}\}$ or in the $\{\ket{+},\ket{-}\}$ basis, either by directly checking that the Holevo-Helstrom conditions hold \cite{Helstrom} or by using the results of \cite{WDA2013} for minimum-cost measurements of symmetric states. One should note, however, that any convex combination of the above projective measurements results in a POVM that gives the same (i.e. the minimum) cost. The minimum cost is $C_{\rm min}=1/8$, as one can see by evaluating the expression
\EQ{C=\sum_{i,j}\frac{1}{4}C_{i,j}\Tr(\Pi_j\rho_i),}
where $\rho_i$ are the BB84 states and $\Pi_j$ are the elements of the POVM used (which are projections, if we are using a projective measurement).
Intuitively, when Bob chooses to measure in the basis which includes the state Alice sent, which happens with probability 1/2, he obtains the correct answer, and thus in this case he never generates a mismatch. When Bob chooses the wrong basis, which happens with probability $1/2$, he causes a mismatch with probability $1/4$. The  overall probability that Bob causes a mismatch is therefore $1/8$.
\end{proof}
The above Lemma means that the probability that Bob generates a mismatch for a single element is at least $C_{\rm min}$, which can be achieved by the above measurement. Thus, in individual attacks, Bob's probability of not being detected for a single element is $(1-C_{\rm min})$. Here it is worth noting that with similar arguments one can compute $C_{\rm max}=3/8$ which is the \emph{maximum} probability of mismatch that one can achieve.

In order to succeed in forging, Bob needs to  correctly declare the part of the signature that Alice sent to Charlie and which Charlie kept. More specifically, he has to avoid mismatches with Charlie's classical signature only for these signature elements. Taking the worst-case scenario, we assume that Bob knows which bits Charlie keeps, before Bob forwards any signature elements to Charlie.
In this case,  for all the elements which Charlie does not keep, Bob can, instead of forwarding the quantum signature element that Alice sent him, send to Charlie a state that will certainly match the declaration that Bob will make later on.
Therefore, for Bob to succeed in forging he must make fewer than $s_v L$ mistakes for the (on average) $L/2$ elements that Charlie received directly from Alice and did not forward to Bob. Taking again the worst-case scenario, we assume that Charlie kept the fewest possible elements, $K=L(1/2-r)$, where $r$ is the abort threshold. Bob can use his own copies of these $K$ elements to make his best guess of a declaration that will be accepted by Charlie, and he is free to perform any measurement that will maximize his probability of forging not being detected.

For now, we will restrict attention to individual attacks. As we showed in Lemma \ref{lemma1}, the probability of mismatch in a single element is at least $C_{\rm min}$.
Bob generates mismatches only for the $K$ elements he needs to guess, while the threshold $s_v$ of accepted mismatches concerns the full signature length $L$. Therefore the effective fraction of mismatches that his guess needs to keep below is $s_v L/K$. For the protocol to be secure we need $C_{\rm min} > s_v L/K \approx 2 s_v$. Then the probability $P({\rm forge} \vert \textrm{ individual attack})$ of ``individual forging'' decays exponentially.  Using the Hoeffding inequalities \cite{Hoeff} for the $K=L(1/2-r)$ elements we obtain the expression
\EQ{P({\rm forge} \vert \textrm{ individual attack})\leq\exp(-2(C_{\rm min}-s_v L/K)^2K).}

{\theorem \label{TH1}The probability that Bob generates a signature that causes fewer than $s_v$ mismatches, if he is only allowed to perform individual measurements, is bounded above by $\exp(-2(C_{\rm min}-s_v L/K)^2K)$, where $K=L(1/2-r)$.}

The same bound also holds for individual adaptive measurements, as the individual states are uncorrelated. This we will show below, as a step of the security proof concerning arbitrary coherent attack strategies.\\

\noindent\emph{Coherent Attacks}:

In the following Lemma we prove that the probability of making a guess that results in a mismatch for (any) $N^{th}$ element cannot decrease, even if Bob applies a joint (coherent) strategy and also post-selects (conditions) on any sequence of outcomes of the previous $(N-1)$ elements. For this, we follow the technique used in the proof by Croke and Kent in \cite{CK2012}. One should note that the following proof applies specifically to the protocol P1, and relies on the particular structure of the BB84 states. Therefore, one cannot immediately generalize this type of proof to other QDS protocols, for instance the ones which use phase-encoded coherent states.

{\lemma\label{lemma2} Suppose Alice generates two copies of a sequence of i.i.d. BB84 states $\ket{\psi_{A_i}}^N_{i=1}$, randomly chosen from the uniform distribution, and gives one copy to Bob and one to Charlie. Charlie makes a USE measurement on each element in his sequence, ruling out one BB84 state $\ket{\psi_{C_j}}^N_{j=1}$ for each element. Bob follows a strategy $S$ and makes a (possibly) 
coherent measurement on his sequence in order to make a guess $\ket{\psi_{B_k}}^N_{k=1}$ for each state.
Let $p_g=p_{A_1,\cdots,A_{N-1}; C_1,\cdots,C_{N-1}; B_1,\cdots, B_{N-1}}$ be the probability that Bob's guess for the $N^{th}$ state that Alice sent ($\ket{\psi_{A_N}}$) is the state that Charlie ruled out ($\ket{\psi_{C_N}}$), conditional on Alice sending the sequence of states $\ket{e_{A_1}},\cdots, \ket{e_{A_{N-1}}}$, Charlie ruling out the states $\ket{e_{C_1}},\cdots, \ket{e_{C_{N-1}}}$ and Bob having guessed the states $\ket{e_{B_1}},\cdots, \ket{e_{B_{N-1}}}$.
Then $p_g\geq C_{\rm min}= 1/8$ for any strategy $S$ and any $\{A_1,\cdots,A_{N-1}; C_1,\cdots,C_{N-1}; B_1,\cdots, B_{N-1}\}$ consistent with $S$.}

\begin{proof}
Suppose some collective strategy $S$ violates this bound for some values $\{A_1,\cdots,A_{N-1},C_1,\cdots,C_{N-1},B_1,\cdots, B_{N-1}\}$. Bob could then proceed in the following way in order to measure a single unknown BB84 state $\ket{\psi_{A_N}}$ of a sequence. Essentially, Bob's strategy below would amount to using the coherent strategy on $N$ qubits, consisting of $N-1$ ``dummy qubits" prepared by himself, and one half of a pair of maximally entangled qubits. If the outcomes at a certain stage in this procedure are as desired, Bob would proceed to ``teleport in" the single unknown BB84 state into the $N^{th}$ place in the qubit sequence, thereby effectively measuring it.
\begin{enumerate}

\item Bob prepares an entangled singlet state of two qubits.

\item Bob prepares $(N-1)$ BB84 states $\ket{e_{A_1}},\cdots,\ket{e_{A_{N-1}}}$ and imagines
that Charlie has (supposedly) ruled out the states $\ket{e_{C_1}},\cdots,\ket{e_{C_{N-1}}}$ which are consistent with the states that Alice (supposedly) sent. We note that Alice and Charlie do not in reality send these states or carry out these measurements. Instead, Bob does everything, in order to use his collective strategy to avoid mismatch for the $N^{th}$ state, which Alice really did send.

\item Bob applies strategy $S$ (ignoring the knowledge of the actual states $\ket{e_{A_1}},\cdots,\ket{e_{A_{N-1}}}$ and the excluded states $\ket{e_{C_1}},\cdots,\ket{e_{C_{N-1}}}$) to the $(N-1)$ BB84 states and one of the entangled qubits.

\item For the first $(N-1)$ states, Bob checks the guesses produced by $S$.

\item If the results do not agree with $\ket{e_{B_1}},\cdots,\ket{e_{B_{N-1}}}$, Bob returns to step 1 with a new singlet and $(N-1)$ new BB84 states. If they do agree, he proceeds to step 6.

\item Bob applies a teleportation operation on the unknown BB84 state $\ket{\psi_{A_N}}$ and the other qubit of the singlet pair and obtains the unitary correction $U=X^aZ^b$. Bob examines the output of the strategy $S$, to see what guess it implies for the $N$th element. Bob applies the corrections $X^aZ^b$ to the classical recorded outcomes.
    By assumption, the adjusted guess
is the state excluded by Charlie with probability $p_g< 1/8= C_{\rm min}$.
\end{enumerate}
This process is bound to proceed to step 6 eventually. The state $\ket{\psi_{A_N}}$ is left isolated until step 6 is reached, and no assumption is made about what state $\ket{\psi_{C_N}}$ Charlie rules out for the $N$th element. Bob therefore has a strategy that produces a guess for a single state $\ket{\psi_{A_N}}$ that happens to be the state
 that Charlie ruled out (thus causing a mismatch), with
 probability $p < 1/8 = C_{\rm min}$, contradicting Lemma \ref{lemma1}.
\end{proof}
Following the same proof one can also prove that no conditional probability can give mismatch probability greater than $C_{\rm max}=3/8$, which is also achieved by an individual strategy. Then, by taking convex combinations of the optimal (maximum-achieving and minimum-achieving) individual measurements one can show that all the probabilities for match or mismatch that one can achieve with conditional measurements, can also be achieved by individual measurements.
For the proof to work, it is crucial that the teleportation correction operations applied to any of the possible states Alice could have sent results in another possible state. This is the case for the BB84 states, but notably it is not the case for two non-orthogonal states.

Using this lemma we can now prove that no coherent strategy can improve Bob's forging probability over the optimal individual attack.
The proof is summarized as follows. First we introduce a modification of the verification procedure, and show that the forging probability using adaptive local measurements in \emph{the modified protocol} upper bounds  the forging probability of any coherent strategy in \emph{the original protocol.}
Following this, we show that local adaptive measurements, in the modified protocol, cannot improve Bob's cheating probability over individual independent measurements. To close the loop, we show that, for individual independent measurements, the cheating probabilities of the original and modified protocol are the same.

In the modification of the verification procedure, Bob selects an order on the qubits (it could be also specified by the protocol), performs local measurements, and declares the states to Charlie sequentially, and finds out, for each state whether the declaration was a match or a mismatch. This allows Bob to modify his local measurements depending on the sequence of previous outcomes, and depending on whether these resulted in matches or mismatches.
Without the loss of generality we will assume that Bob will measure his qubits (i.e. the corresponding ancilla states) in the natural order of the indices.

Let $V$ denote the a forging strategy, using coherent strategies, on the original protocol.
The overall forging probability can always be written as
\EQ{p({\rm forge})=\sum_{\sigma\in A}p(\sigma)}
where $\sigma=(x_1,\cdots,x_L)$ is any string of matches/mismatches (say, the variable $x_i\in\{0,1\}$ denotes whether the $i^{th}$ declared element matches ``$0$'' or mismatches ``$1$'' the excluded element of Charlie) and $A$ is the set of all strings $\sigma$ that have fewer than $s_v L$ 1's (mismatches).

Then, for the strategy $V$, each probability $p_{V}(\sigma)$ of the individual event $\sigma = (x_1,\cdots,x_L)$ can be written using the chain rule as
\EQ{
p_{V}(\sigma) = p_{V}(x_1,\cdots,x_L) = p_{V}(x_1)p_{V}(x_2 \vert x_1) p_{V}(x_3 \vert x_2,x_1) \cdots p_{V}(x_L \vert x_{L-1}, \ldots x_1)
}
By Lemma \ref{lemma2} and the comment thereafter regarding the maximal probability of causing a \emph{mis}match, we have that
\EQ{
C_{\rm max}\geq p_{V}(x_k = 1 \vert x_k, x_{k-1},\ldots x_2,x_1) \geq C_{\rm min}, \forall k,\ \forall (x_k, x_{k-1},\ldots x_2,x_1)}
where both $C_{\rm min}$ and $C_{\rm max}$ can be achieved by local
measurement strategies.
This implies that for each sequence of prior outcomes $(x_k, x_{k-1},\ldots x_2,x_1)$, there exists a local strategy/measurement $M_{k,  x_{k-1},\ldots x_2,x_1}$ acting only on qubit $k$, which is a convex combination of the strategies maximizing a mismatch and a match such that
\EQ{
p_{M_{k,  x_{k-1},\ldots x_2,x_1}} (x_k) = p_{V}(x_k \vert x_k, x_{k-1},\ldots x_2,x_1).
}

This proves that for every coherent strategy in the original protocol, there exists  an adaptive local strategy in the modified protocol which recovers the probability distribution over matches/mismatches of the coherent strategy.
Hence we have:
\EQ{
P({\rm forge} \vert \textup{coherent\ attack}, \textup{original\ protocol}) \leq P({\rm forge} \vert \textup{adaptive\ local\ attack},  \textup{modified\ protocol}).}

Next, we show that the best adaptive local strategy in the modified protocol is the optimal individual (non-adaptive) strategy.
To see this, first note that at the $k^{th}$ step of the verification procedure, since all the measurements made so far have been local, the remainder of the $L-k$ qubits have not been perturbed. This implies that the probability of obtaining a match on the next, $(L-k)^{th}$ qubit, does not depend on the previous $k$ measurement outcomes (or declarations of match/mismatch), since the qubit states are not correlated (and neither are the verification measurements of Charlie).
This intuitively shows that the optimal strategy are local optimal measurements, but for completeness, we prove this formally.
First we give a trivial claim: given $L'$ signature states and some threshold $k$, the probability $P(X_{match} \geq k)$ of getting at least $k$ matches is higher than or equal to the probability $P(X_{match} \geq k+1)$ of getting at least $k+1$ matches, that is $P(X_{match} \geq k) \geq P(X_{match} \geq k+1)$. This is trivial as the event $X_{match} \geq k$ is contained in the event $X_{match} \geq k+1$.
In the remainder we will use $k$ to denote the forging threshold of matches so $k = \lceil L - s_v L \rceil.$
Suppose Bob is at some stage $l$ of the (modified) verification procedure. There are two possibilities, either he obtained the $k$ required matches for cheating or he did not.
If he obtained the matches, then the remainder of declarations does not change his (unit) forging probability, and any strategy (in particular, the optimal local (non-adaptive) strategy ) is optimal.
Alternatively, he still needs to obtain $k' \leq k$ matches on the remainder of $L-l$ qubits.
His forging probability, at that point, is given by
\EQ{
p({\rm forge}) = p(x_l = 0)P(X_{match} \geq k'-1) + p(x_l = 1)P(X_{match} \geq k').
}
That is, either he gets the $l^{th}$ qubit correctly, after which he needs only $k'-1$ matches, or he does not, and still requires $k'$ matches for the remaining qubits.
Since  $P(X_{match} \geq k'-1) \geq P(X_{match} \geq k')$, the expression above is optimized by maximizing $p(x_l = 0),$ which occurs with the optimal local measurement (and is independent of any previous outcomes).
Since this argument holds for all $l,$ this means that the local (non-adaptive) strategy is optimal, \emph{i.e.}
\EQ{
 P({\rm forge} \vert \textup{adaptive\ local\ attack},  \textup{modified\ protocol}) \leq
 P({\rm forge} \vert \textup{individual\ attack},  \textup{modified\ protocol}).
 }
However, since the optimal individual strategy does not use the declarations Charlie provides in the modified verification protocol, this implies that the forging probability using non-adaptive measurements in the original  protocol, and in the modified protocol, are equal:
\EQ{
 P({\rm forge} \vert \textup{individual\ attack},  \textup{modified\ protocol}) = P({\rm forge} \vert \textup{individual\ attack},  \textup{original\ protocol}),
}
and since coherent attacks contain individual attacks, we have
\EQ{
P({\rm forge} \vert \textup{coherent\ attack}, \textup{original\ protocol})  = P({\rm forge} \vert \textup{individual\ attack}, \textup{original\ protocol}).
}

Combining this result with Theorem \ref{TH1},  we have proven the following main Theorem:

{\theorem The probability that Bob, by measuring his sequence of states, generates a signature declaration with fewer than $s_v$ mismatches is bounded by $p({\rm forge})\leq\exp(-2(C_{\rm min}-s_v L/K)^2K)$, where $K=L(1/2-r)$. That is, the forging probability of the presented QDS protocol decays exponentially with the signature length $L$ for all possible attacks.}\\

\noindent\emph{Further remarks}:

We had previously stressed that the presented proof of security against general (coherent) forging attacks crucially depends on the choice of BB84 states for the signature elements.
In particular, step 6 of Lemma \ref{lemma2} fails in the general case.
The results of Lemma \ref{lemma2} can be extended to any set of states $\mathcal{S}$ for which there exists a teleportation procedure with correction operators which leave the set $\mathcal{S}$ invariant.  This will imply that the correction operators simply permute the input set, which allows for Bob to `correct' the classical outcome of his post-selected strategy which would violate the individual measurement bound.
In particular, Lemma \ref{lemma2} does not hold for the so-called B92 states $\{\ket{0},\ket{+}\}$. One can construct a counter-example with just two copies of B92 states (see example in \cite{WDA2013}). Alice sends one of the four states $\{\ket{00},\ket{0+},\ket{+0},\ket{++}\}$. Let Bob measure in the basis
    \EQ{\ket{\phi_{++}}&=&1/\sqrt2(\ket{01}+\ket{10})\\
\ket{\phi_{+0}}&=&1/\sqrt2(\ket{0-}+\ket{1+})\\
\ket{\phi_{0+}}&=&1/\sqrt2(\ket{+1}+\ket{-0})\\
\ket{\phi_{00}}&=&1/\sqrt2(\ket{+-}+\ket{-+}),}
and make the relevant declaration to get a probability distribution $p_M(x_1,x_2)$ on matches and mismatches. This measurement by construction guarantees that he never gets both elements wrong (e.g. if he obtains $\phi_{++}$, it means that Alice did not send the state $\phi_{00}$, so he obtains at most one mismatch). Now conditioning on the first qubit to be a mismatch, he obtains $p_M(x_2=0|x_1=1)=1$. This is clearly better than the optimal local strategy, which can never succeed with unit probability, $1=p_M(x_2=0|x_1=1)>p_{local}(x_2=0)$.

When the states used for QDS are not suitable for the type of security proof we presented here, following \cite{CK2012} one could suggest an alternative proof based on maximum confidence measurements (MCM)\cite{CAB2006}. The basic idea here would be to produce a bound by considering MCM's, and further conditioning on these always producing a conclusive outcome. Due to this post-selection the obtained bound is not tight, but can be applied to a larger variety of quantum states. However, this approach still cannot be applied to linearly independent states (such as the B92 states, or phase-encoded coherent states), as in this case it yields a trivial bound of $p_{forge}=1$.

\subsection{Security of protocol P2}
We will first show for protocol P2 that the probability for repudiation decreases exponentially with the length $L$ and then do the same for the forging probability.
\subsubsection{Security against repudiation}

In order to repudiate, Alice must make Bob accept the message while Charlie rejects it. Since Bob has to accept the message, Alice's declaration must agree with all the elements of $PrivKeyB_m$. On the other hand, for Charlie to reject the message, he needs to detect at least $s_vL$ mistakes. These should all come from $PrivKeyC_m$. Coming back to the requirement that Bob has to accept the message, we see that none of the elements that Bob receives from Charlie should include a mismatch. Since Charlie sends each bit of his $PrivKeyC_m$ to Bob with probability $1/2$, then
if there are $R$ mismatches in $PrivKeyC_m$, the probability for Bob to see no mismatches is $(1/2)^{R}$. It is also clear that the best strategy for Alice is to send exactly $R=s_vL$ mismatches to Charlie, and this leads to Alice's optimum repudiation probability
\EQ{p(rep)\leq (1/2)^{s_vL}}
which decays exponentially as the length of the signature $L$ increases.

\subsubsection{Security against forging}

Bob, in order to forge, must give a declaration that has fewer than $s_vL$ mismatches. Note that this protocol is essentially classical, so if Alice sends a bit that does not agree with her future declaration, then the recipient detects the mismatch deterministically. If Charlie sends more than $L(1/2+r)$ bits of his private key to Bob, then the protocol is aborted by step 4 of the distribution stage. We assume the worst-case scenario (for the honest participants, Charlie and Alice) that Charlie has sent exactly $L(1/2+r)$ elements of his private key to
Bob.
This means that Bob must guess the remaining $K=L(1/2-r)$ bits in $PrivKeyC_m$, making fewer than $s_vL$ mistakes. The expected probability of error for a single guess is $1/2$. The empirical mean number of wrong guesses $\bar X$ needs to be
less than $s_vL/K$ (in other words, Bob should make fewer than $s_vL$ mistakes among the $K$ elements he is required to guess).
This, using Hoeffding's inequalities \cite{Hoeff}, implies that the probability to forge is bounded by
\EQ{p({\rm forge})=p(1/2-\bar X\geq 1/2-s_vL/K)&\leq& \exp \left[-2(1/2-s_vL/K)^2K\right]\nonumber\\
&=&\exp\left[-4\left(1/4-\frac{s_v}{1-2r}\right)^2L(1-2r)\right].}
which, provided that $s_v<1/4-r/2$, decays exponentially as $L$ increases. Since typically $r$ will be chosen to be small, this condition agrees with the intuitive picture. A forger will on average guess half of the elements
correctly, so he would typically make $O(L/4)$ mistakes. Therefore, choosing $s_v$ smaller than $1/4$ guarantees the security.

Finally, it is important to note, that unlike in protocol P1 and its variants, in protocol P2 Alice sends \emph{different} signatures to Bob and Charlie. If Alice was to send the same signature to Bob and Charlie, and they are aware of this, then forging would be possible.


\begin{thebibliography}{99}

\bibitem{Rivest1990} R. L. Rivest, Handbook of Theoretical Computer Science, p. 717-755, Elsevier (1990).

\bibitem{QDS}
D. Gottesman and I.  Chuang, 
{arXiv:quant-ph}/0105032v2 (2001).

\bibitem{ErikaOrig}
E. Andersson, M. Curty, and I. Jex, 
{Phys. Rev. A} {\bf 74}, 022304  (2006).

\bibitem{OurNatComm}
P. J. Clarke {\it et al.}, 
{Nat. Commun.} {\bf 3}, 1174 (2012).

\bibitem{QDS_prl} V. Dunjko, P. Wallden, and E. Andersson, 
Phys. Rev. Lett. \textbf{112}, 040502 (2014).

\bibitem{QDS_exp} R. J. Collins, R. J. Donaldson, V. Dunjko, P. Wallden, P. J. Clarke, E. Andersson, J. Jeffers, and G. S. Buller,  Phys. Rev. Lett. \textbf{113}, 040502 (2014).

\bibitem{Lamport} L. Lamport, Technical Report CSL-98, SRI International (1979).

\bibitem{Hanaoka2000} G. Hanaoka, J. Shikata, Y. Zheng and H. Imai,
Advances in Cryptology, LNCS, \textbf{1976} 130, (2000).

\bibitem{SwSt2011} C. Swanson and D. Stinson, 
in [Information Theoretic Security] , First Edition, S. Fehr, Ed., Springer, Berlin Heidelberg, 100-116 (2011).

\bibitem{CR1991} D. Chaum and S. Roijakkers, 
in Proceedings of the 10th Annual International Cryptology Conference on Advances in Cryptology, A. Menezes and S. A. Vanstone, Eds., 206–
214 (1991).



\bibitem{MAC} M. N. Wegman and J. L. Carter, Journal of Computer and System Sciences, {\bf22}, 265 (1981).


\bibitem{BB84} C. H. Bennett and G. Brassard, in Proceedings of the IEEE International Conference on Computers, Systems, and Signal Processing, Bangalore, p. 175 (1984).

\bibitem{stevebook}
S. Barnett, {\it Quantum Information}, Oxford University Press, pp 103-104 (2009).

\bibitem{OppenUSE}
S. Bandyopadhyay, R. Jain, J. Oppenheim, and C. Perry, 
Phys. Rev. A {\bf 89}, 022336 (2014). 

\bibitem{WDA2013} P. Wallden, V. Dunjko and E. Andersson, J. Phys. A {\bf 47}, 125303 (2014).

\bibitem{CK2012} S. Croke and A. Kent, Phys. Rev. A {\bf 86}, 052309 (2012).

\bibitem{Kent2012} A. Kent, Phys. Rev. Lett. {\bf 109}, 130501 (2012).

\bibitem{bcexpt} T. Lunghi et al, Phys. Rev. Lett. {\bf 111}, 180504
  (2013).

\bibitem{idquantique} id Quantique, http://www.idquantique.com, visited 21/03/2014.

\bibitem{magiq} MagiQ Technologies, http://www.magiqtech.com, visited 21/03/2014.

\bibitem{quintessence} QuintessenceLabs, http://quintessencelabs.com/, visited 21/03/2014.

\bibitem{sequrenet} SeQureNet, http://www.sequrenet.com, visited 21/03/2014.

\bibitem{anhui} Quantum Communication Technology Co. Ltd. Anhui,  http://www.quantum-info.com/en.php, visited 21/03/2014.

\bibitem{AQM2002}
H. Barnum, C. Cr{\'e}peau, D. Gottesman, A. Smith and A. Tapp
\newblock{{\em Proceedings of the 43rd Symposium on Foundations of Computer Science}}, 2002.

\bibitem{Renner05} R. Renner, PhD Thesis, preprint quant-ph/0512258, 2005.

\bibitem{Renner07} R. Renner, Nat. Phys. {\bf 3}, 645, 2007.

\bibitem{Hoeff}
 W. Hoeffding, Probability inequalities for sums of bounded random variables.
 {\em J. Amer. Statist. Assoc.},
  {\bf 58} 301, (1963).

\bibitem{Helstrom}
C. W. Helstrom
\newblock {\em { Quantum detection and estimation theory}}.
\newblock Academic Press, New York, 1976.


\bibitem{CAB2006} S. Croke, E. Andersson, S. Barnett, C. Gilson and J. Jeffers, Phys. Rev. Lett. \textbf{96}, 070401 (2006).

\end{thebibliography}
 \end{document}